\documentclass[runningheads]{llncs}
\usepackage{graphicx}
\usepackage{wrapfig}
\begin{document}
\title{Materials Science Ontology Design with an Analytico-Synthetic Facet Analysis Framework\thanks{This work is supported by NSF-OAC 2118201, NSF-IIS 1815256.}}

\author{Jane Greenberg\inst{1} \and
Scott McClellan\inst{1} \and
Xintong Zhao\inst{1} \and
Elijah J Kellner\inst{2} \and
David Venator\inst{3}\and
Haoran Zhao\inst{1}\and
Jiacheng Shen\inst{1}\and
Xiaohua Hu\inst{1}\and
Yuan An\inst{1}}

\institute
{
  \inst{}%
  Metadata Research Center\\
  Drexel University
  \and
  \inst{}%
  College of Science and Engineering\\
  Winona State University
  \and
  \inst{}
  McCormick School of Engineering\\
  Northwestern University
}

\authorrunning{Greenberg et al.}
\maketitle              % typeset the header of the contribution
\begin{abstract}
Researchers across nearly every discipline seek to leverage ontologies for knowledge discovery and computational tasks; yet, the number of machine readable materials science ontologies is limited. The work presented in this paper explores the Processing, Structure, Properties and Performance (PSPP) framework for accelerating the development of materials science ontologies. We pursue a case study framed by the creation of an Aerogel ontology and a Battery Cathode ontology and demonstrate the Helping Interdisciplinary Vocabulary Engineer for Materials Science (HIVE4MAT) as a proof of concept showing PSPP relationships. The paper includes background context covering materials science, the PSPP framework, and faceted analysis for ontologies. We report our research objectives, methods, research procedures, and results. The findings indicate that the PSPP framework offers a rubric that may help guide and potentially accelerate ontology development.

\end{abstract}
\section{Introduction}
An ontology is an explicit language that includes a vocabulary. An ontology, and hence the underlying vocabulary, frequently represents a domain or area of knowledge. Machine readable ontologies are made explicit by adhering to grammatical rules and through the application of enabling technologies, such as the the Resource Description Framework (RDF) and Web Ontology Language (OWL). These technologies have advanced the development of ontologies, including the transformation of analog disciplinary vocabularies into machine readable ontological structures. Furthermore, they enable ontologies to support information discovery, reasoning, data interoperability, linked data applications, and other computational tasks. Today hundreds of ontologies in biomedicine, biology, and related disciplines are accessible via portals, such as the National Center for Biomedical Ontology’s Bioportal \cite{NCBO} and the OBO Ontology register\cite{OBO}--both of which have been operational since the early and mid-2000’s. In comparison, the number of machine readable materials science ontologies is limited, with MatPortal\cite{matportal} hosting approximately 25 ontologies. The limited number of materials-science ontologies challenges researchers to identify approaches for accelerating their development.

The research presented in this paper seeks to address this challenge by exploring analytico-synthetic faceted classification\cite{b21} as a framework for accelerating the development of materials science ontologies. In pursing this task, it is important to examine why there are a relatively small number of materials science ontologies. One likely factor is R\&D (research and development) funding allocations with the life-sciences, which includes medicine and biology, has generally exceeded other disciplines\cite{b22}. A more obvious factor is the broad context that materials science encompasses, spanning engineering, physics, chemistry, and other interconnected disciplines. Finding a unifying approach to any knowledge-based task is difficult across any interdisciplinary and transdisciplinary field, and materials science is among one of the more extensive disciplines in its topical and disciplinary span. A final factor to note here is the absence of a clear underlying semantic framework. While materials scientists may examine biomedical ontologies for guidance and research  support, ontological structures from these disciplines may not resonate or sufficiently to inform development of material science ontologies. This predicament motivates us to examine the Processing, Structure, Properties and Performance (PSPP) framework for ontology development. The work presented in this paper specifically investigates the application of the PSPP model for developing materials science ontologies. We present a case study framed by the creation of two exploratory ontologies, one focused on Aerogels and another on Cathode Batteries. Additionally, we demonstrate the use of these ontologies with the Helping Interdisciplinary Vocabulary Engineer for Materials Science (HIVE4MAT) indexing application to illustrate PSPP relationships extracted materials science literature. The following sections include background context, the guiding objectives, methods, research procedures, and results.  The last two sections include a discussion and a conclusion, which underscores key takeaways and next steps.

\begin{wrapfigure}{R}{0.6\textwidth}
\centering
\includegraphics[width=0.55\textwidth]{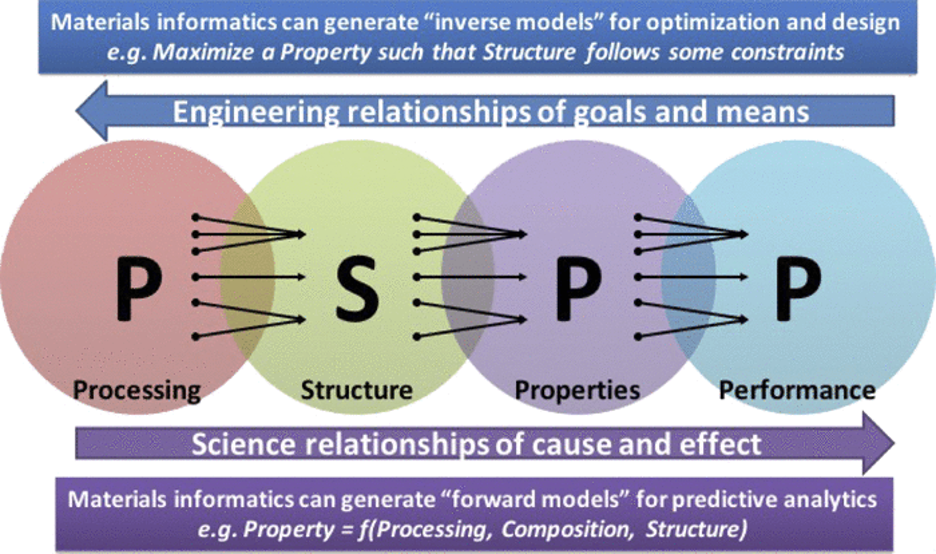}
\caption{\label{fig1}Materials Paradigm showing the scientific and engineering approaches to the field \cite{b10}}
\end{wrapfigure}

\section{Background and Motivation}

\subsection{Materials Science and PSPP}
In exploring the PSPP framework, it is important to understand the expanse and context of materials science. Materials science is an interdisciplinary field that utilizes physics, chemistry, and engineering principles to identify new materials and improve existing ones. The field focuses on solids, which have a defined structure unlike liquids or gases. These defined structures give rise to specific properties, and understanding how and why certain structures correspond to useful properties is central to the field. There are four general classes of materials – metals, polymers, ceramics, and composites – each with a different type of structure that make the class more favorable for certain applications. In each field there are infinite potential materials, each with a unique set of properties.

At a very general level, materials science research follows two pathways. The historical and still significant path is method-driven. Research following this pathway focuses on the scientific relationships between materials processing and performance. This approach allows a researcher to gain a thorough understanding of how certain processes lead to certain structures, which give rise to unique properties, and allows a material  to perform in specific ways. This has long been achieved through experimentation. As understanding of these relationships across processing, structure, properties, and performance has grown, an inverse research pathway has emerged, and has become crucial to material-related engineering. The goal of this other pathway is to find a material and materials processing method that achieves the desired material performance. This method has advanced as researchers pursue data-driven approaches, and allow researchers to more efficiently simulate processing in order to measure performance. This framework is illustrated in Figure \ref{fig1}. 

This framework may inform the development of materials science ontologies, which are increasingly important for data-driven research including AI. The next section discusses facet analysis and how this approach connects with ontology development.

\subsection{Facet Analysis and Ontologies}

The fundamental approach to facet analysis and organization has been practiced since ancient times. Facets are defined as mutually exclusive categories. The Pinakes, the catalog Callimachus (310/305–240 BCE) created for the Library of Alexandria, provided disciplinary classes, which knowledge organization researchers have discussed as topical facets. Supporting this view is the Aristotelian notion of classification, where categories are determined by necessary and sufficient conditions of their existence\cite{b14}. As direct evidence is scarce, knowledge of these developments descended through the writings of Pliny the Elder and other classical writers who studied at the library.

Faceted knowledge organization systems gained popularity in the late 19th and early 20th  century schemes, such as the Dewey Decimal System’s use of tables for geographic regions, languages, and other areas \cite{b13,b18}; Bliss’s methodological organization of agents, operations, properties, materials, processes, parts, types, and thing \cite{b12}; and, Colon Classification System \cite{b19}, perhaps the most recently referenced classification system, developed by the mathematician/librarian Shialy Rammarita Ranganathan. Colon classification is based on postulates representing five fundamental categories, each an isolated facet, with five Fundamental Categories (FC) which include: Personality, Matter, Energy, Space and Time (PMEST).

In all of these examples, knowledge representation follows an approach known as analytico-synthetic classification, where a class notation is constructed through relationships connecting the facts. Faceted approaches have an advantage over hierarchical systems, the latter of which tend to isolate related topics and sub-topics in a taxonomic way that can lead to overlapping and redundant representations with the result that a user can miss relevant information having to in initiate their search at a lower than intended level \cite{b2}. While faceted systems frequently have an element of hierarchy, the level plane for entry enables a more flexible, accommodating approach, and one that is worth exploring in materials science as viewed in the work of \cite{b6}. In sum, the guiding principles underlying faceted systems and the significance of the PSPP framework motivate further exploration of the analytico-synthetic approach for materials science ontologies, and shape the objectives that have guided our work.

\section{Research Goals and Objectives}
Our overall goal is to investigate the PSPP framework as a method for streamlining the development of materials science ontologies. The individual component of PSPP provides a top-level scaffold of facets and can support analytico-synthetic classification, leading to discovery of new knowledge. We pursued this work by developing two materials science ontologies, one focused Aerogels, which are low-density, open-pored nanostructured materials that support low thermal conductivity, and high adsorption capacity (Feng et al. 2020), and the other focused on on Battery Cathode materials, as lithium-ion batteries have rapidly become the standard for many applications\cite{b16}. Specific objectives guiding our work include:

1. Exploring literature covering these two areas

2. Developing two baseline/preliminary ontologies, one for aerogels and one for lithium-ion battery cathode materials 

3. Demonstrating the application of the ontologies for displaying relationships and knowledge discovery, using the HIVE4MAT ontology application\cite{b20}.

\section{Methods  and Research Design  Procedures}
We pursued the above objectives using a mixed methods approach combining the case study approach, card sorting, and demonstration. The case method, drawing from classic work of Fidel\cite{b23}, allows researchers to explore a topic at a more indepth level as we explicitly defined the ontology space covering Aerogels and Battery Cathode materials using the PSPP framework. The card-sorting method informed this process, as we grouped terms according to the PSPP framework, under the facets processing, structure, properties, and performance. Finally, we demonstrated the application of both ontologies for knowledge extraction, using the HIVE4MAT ontology application. This offered a proof of concept.

Our procedures involves following seven steps:

\begin{itemize}
    \item \textbf{Step 1: Locate and confirm an area of study}. We started with broader materials science categories which were then narrowed to increase specificity of the vocabulary. The initial categories of gels and batteries were viewed as overly broad, which led to the selection of battery cathodes and aerogels which provide a balance of specificity and breadth. It was important to confirm that the area of study had a definable aspect of materials science and yielded a  manageable, useful set of terms for the ontology.
    \item \textbf{Step 2: Term collection}. Term collection was pursued using both automatic and manual review of relevant literature. Terms were selected that related to the two  defined topical areas, Aerogels and Battery Cathode materials.
    \item \textbf{Step 3: Term sorting}. Terms were sorted into relevant PSPP facets based upon common features. Following the initial faceted analysis, terms were arranged  hierarchically, defined by “is-a”relationship between parent and child concepts. 
    \item \textbf{Step 4: Establishing Relationships}: The process of defining relationships between terms as introduced as the sorting activity, and exclusively focused when the sorting was completed.
    \item \textbf{Step 5: Ontology Encoding}: We used WebProtege to visualize the hierarchy and code the terms and their relationships, resulting in the two example ontologies. During this process, some ontology refinement also took place, discarding extraneous terms from each ontology, and the terms that remained were further scoped. During this step, we were able to also examine several symmetrical relationship functions, including their domains and ranges. These qualifications of the relationships and concepts reinforced the interconnections among the categories. 
    \item \textbf{Step 6: HIVE4MAT integration}: The baseline Aerogel Ontology and the Battery Cathode Ontology were both integrated into the HIVE4MAT ontology server, and general functionalities of search, browse and index were confirmed.
    \item \textbf{Step 7: Demonstration}: The final step in ontology development was deployment and testing in HIVE4MAT, which offered a proof of concept.
\end{itemize}

\section{Results}
Results reported here cover the ontology structures, their display in Protege, and HIVE4MAT outputs.

\subsubsection{Aerogel Ontology}
\begin{figure}
\centering
\includegraphics[width=0.3\textwidth]{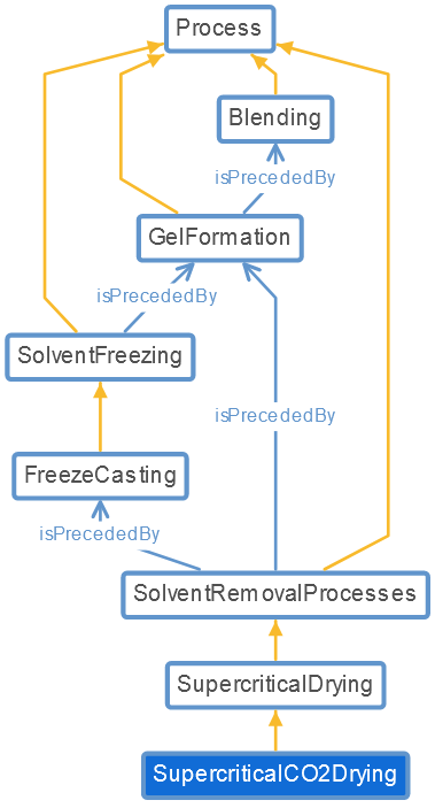}
\caption{Aerogel Ontology: Entity and Relationship Display} 
\label{fig2}
\end{figure}

This aerogel ontology contains terms that are related to the processing, structure, properties, and performance of aerogels. It should be noted that this is only a baseline ontology and that many terms relating to aerogels, especially newly developed terms, are likely not contained within the ontology and that future updating and expansion of the ontology is recommended. This case study ontology contains 7 performance terms, 38 processing terms, 30 property terms, and 32 structure terms, for a total of 107 terms. These counts include both the preferred label and any synonyms for the listed concepts. The aerogel ontology also contains 4 relationships, which are \textit{isSynthesizedBy}, \textit{isDependentOn}, \textit{isDerivedFrom}, and \textit{isPrecededBy}. The \textit{isSynthesizedBy} relationship connects structures with the process used to synthesize them; the \textit{isDependentOn} relationship connects property terms to performance terms that are based on them; the \textit{isDerivedFrom} relationship connects performance terms to the property terms that they are mathematically derived from; and the \textit{isPrecededBy} relationship connects process terms in the order that they occur in a standard synthesis or procedure (Figure \ref{fig2}).

\begin{figure}
\centering
\includegraphics[width=0.5\textwidth]{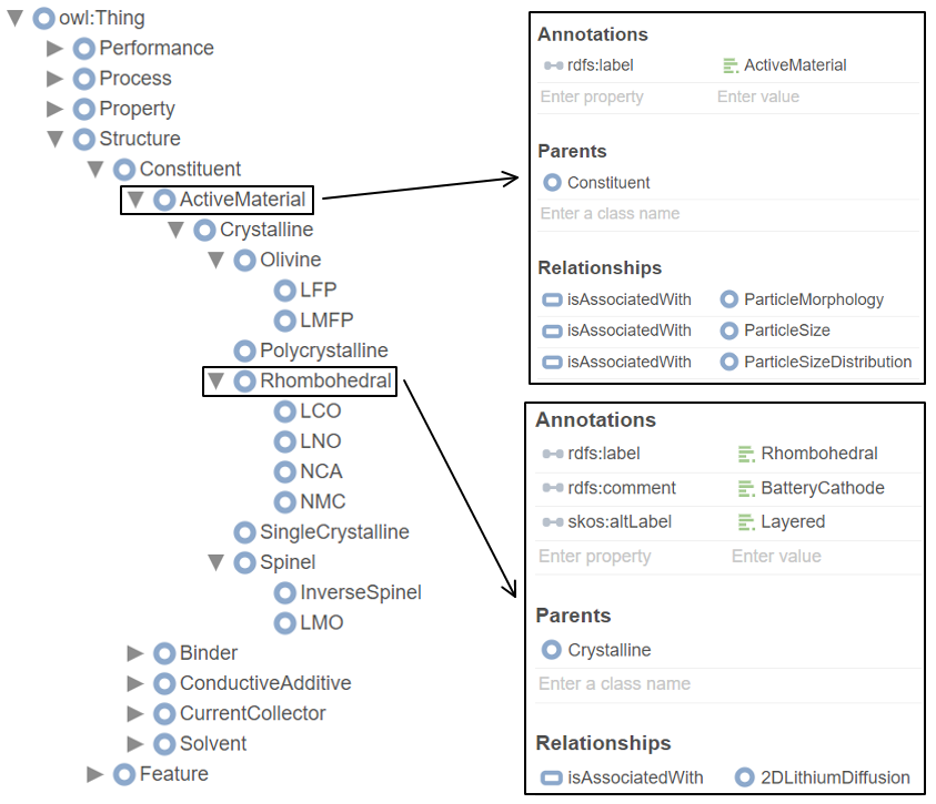}
\caption{Battery Cathode Ontology: Protege Faceted and Hierarchical Structure} 
\label{fig3}
\end{figure}

\subsubsection{Battery Cathode Ontology}
As an area of high research interest, there are a wealth of terms associated with the study of lithium-ion battery cathodes. This ontology serves as a baseline, capturing many of the foundational terms involved in the processing, structure, properties, and performance of cathode materials. This ontology is not comprehensive; with studies focusing on certain aspects of the cathode (introducing novel processing methods or using constituents of novel composition) likely to have relevant terms not captured in this ontology. In order to truly capture all terms in this space, the process would likely have to be automated. 

This ontology contains 153 independent terms, exclusive of alternative labels, across each part of the materials science paradigm. There are 20  processing terms, 58 structure terms, 39 property terms, and 36 performance terms. These terms are organized hierarchically, as shown in the figure below. Relationships are also used to connect  terms among separate branches. One example, as shown in the figure below, is relating certain structure terms (particle size/size distribution/morphology) to the part of the composite cathode that they relate to (the active material) through the relationship \textit{isAssociatedWith}. Other relationships include \textit{isPreceededby}, to connected processes in the general synthesis procedure, and \textit{isDerivedFrom}, to connect performance metrics to the properties they are mathematically derived from (Figure \ref{fig3}).

\begin{figure}
\centering
\includegraphics[width=\textwidth]{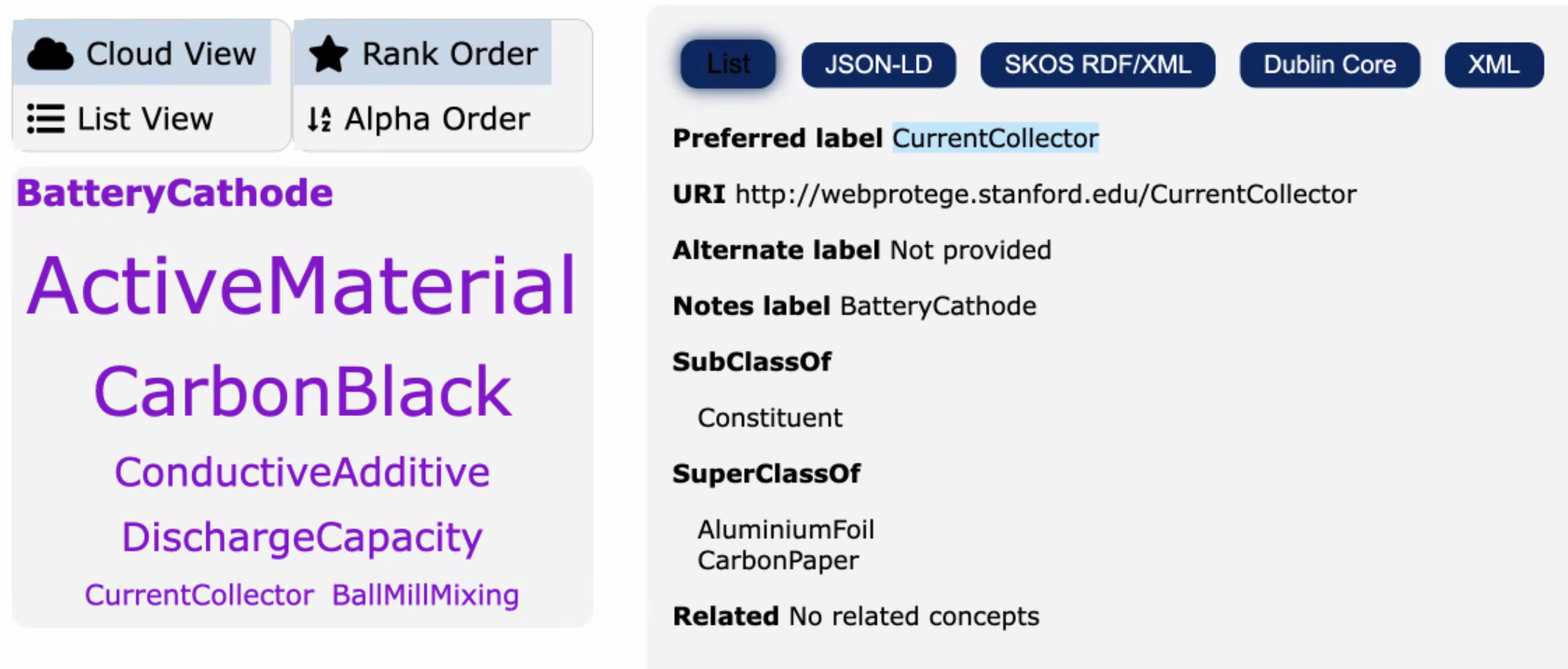}
\caption{HIVE4MAT: Knowledge Extraction Demonstration with Battery Cathode Ontology} 
\label{fig4}
\end{figure}

\begin{figure}
\centering
\includegraphics[width=0.9\textwidth]{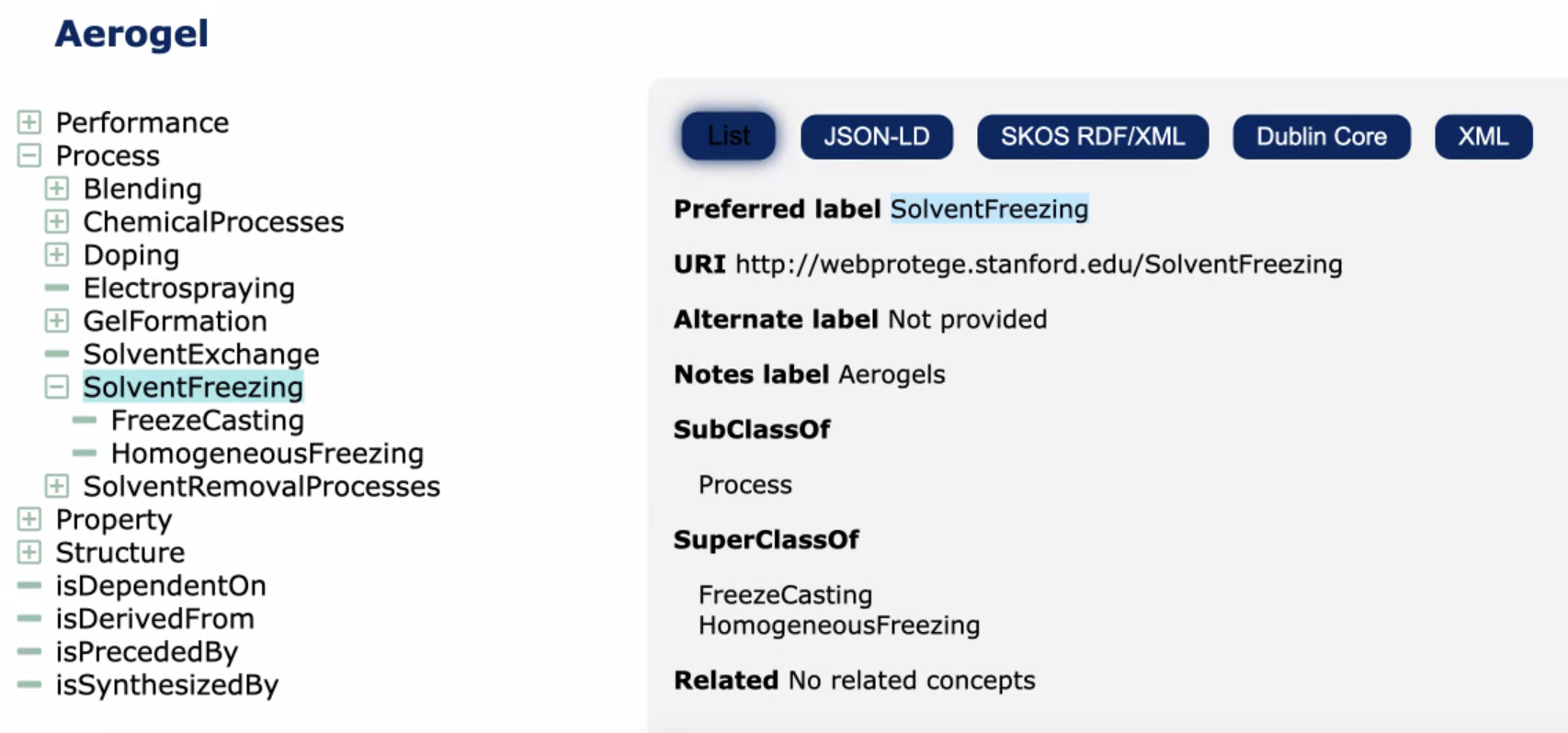}
\caption{HIVE4MAT: Browse display for Aerogel Ontology} 
\label{fig5}
\end{figure}

Finally, we tested the functionality of the ontologies in the HIVE4MAT ontology server. Figures 4 and 5 show that the ontologies are browsable and support knowledge extraction when tested with unstructured scholarly big data. The work with HIVE4MAT offers a proof of concept. Figures 4-5 demonstrate the use of HIVE4MAT. Figure 4 presents knowledge extraction results (left-hand side) in using the Battery Cathode Ontology in HIVE4MAT. The right-hand side includes the term display for "CurrentCollector" and associated ontological relationships. Figure 5 presents the the HIVE4MAT browse display for Aerogel Ontology which situates a term, "SolventFreezing," within the hierarchy of terms and relationships of the Aerogel ontology.

\section{Discussion}

The case study presents a proof of concept using the PSPP framework for the  Aerogel and Battery Cathode ontologies. The case study also demonstrates the HIVE4MAT application for knowledge extraction and the ontology browse feature. The narrowed scope of the two ontologies allowed them to extract specific topical aspects of documents. Moreover, the narrow scope underlying the exploratory ontology construction largely falls in line with contemporary methods which focus on area-specific topics and link to broader domain- or upper-level ontologies for general terms. However, the limited scope of the vocabularies similarly narrows the applications of the ontologies to smaller tranches of materials science and possibly the relationships between and among concepts. These limitations are partly the result of decreased time and human resources; most ontologies are the result of subject matter experts reaching consensus about terminology and relationships, sometimes over the course of several years, such as the Elemental Multiperspective Materials Ontology (EMMO), which is an ongoing initiative since 2016 (https://github.com/emmo-repo/EMMO/). It's also important to point out that the ontologies developed for exploring faceted classification have an extensive genealogy stemming primarily from \cite{b24} which describes a set of core ontologies for materials science based around the following aspects: substance, process, property, and environment. In addition, the ontologies postulated in this paper align closer with early efforts such as the Plinius ontology of ceramic materials \cite{b25} for using ontologies for automatic classification of materials science literature. The use of HIVE4MAT broadens the graph of possible ontology and vocabulary classification structures to show possible unknown connections between smaller test ontologies such as our Aerogel or Battery Cathode as well as those such as EMMO (to be tested at a future date).

The exploration of facet analysis for material science ontology construction proved a useful heuristic for the constructors and deciding how to categorize terms. The method sparked interesting discussions about how to classify terms which seemed to fall into several facets. The Performance facet proved the most complex to handle as it attempts to represent a more dynamic aspect of a material or process. Relationships between the facets proved somewhat complex to model given their more dynamic nature. The discussions regarding relationships offer insight into future HIVE4MAT improvements, including future support for OWL relationships beyond \textit{isA} and \textit{hasA}. Finally, this discussion and earlier efforts in the faceted ontology space demonstrate this topic warrants further exploration.

%%%Finally, it's important to point out that the ontologies developed for exploring faceted classification have an extensive genealogy stemming primarily from \cite{b24} which describes a set of core ontologies for  materials science based around the following: substance, process, property, and environment, Moreover, the ontologies postulated in this paper align closer with early efforts such as the Plinius ontology of ceramic materials \cite{b25} for using ontologies for automatic classification of materials science literature. The use of HIVE4MAT broadens the graph of possible ontology and vocabulary classification structures to show possible unknown connections between smaller test ontologies such as our Aerogel or Battery Cathode as well as those such as EMMO (to be tested at a future date). In this respect, the PSPP facet model allows for a simpler, more streamlined method of producing local ontologies in an attempt to decrease the overhead associated with consensus building.

\section{Conclusion}
This paper reports on research investigating analytico-synthetic faceted analysis for building material science ontologies. The case study approach, along with the card sorting exercise,  supported the construction of two base-level ontologies following the PSPP framework, and we demonstrated the application of both ontologies for knowledge extraction using the HIVE4MAT ontology application. 

Ontology development in any area generally involves a team of multiple people and is an interactive process, with refinements over time. Reaching consensus among team members often proves problematic due to disparate theoretical or methodological stances of the participants. Additionally, terms are pruned and added as knowledge changes and new term relationships evolve. The work presented here provides a foundation for further exploration of PSPP facets for ontology development. The framework proved useful overall in guiding the term sorting and overall ontology construction; although, not all of the facets had equal representation. Future work will address issues surrounding the distribution of terms across the facets and how the relationships can be better deployed and displayed as part of an automatic indexing pipeline supporting knowledge extraction. 
\section{Acknowledgment}
This work is supported by NSF-OAC 2118201, NSF-IIS 1815256.


\begin{thebibliography}{8}
\bibitem{b1}
Sciences, Medicine, others (2019). frontiers of materials research: A decadal survey. National Academies Press.

\bibitem{b2}
Ghosh, S., Panigrahi, P. (2015). Use of Ranganathan’s analytico-synthetic approach in developing a domain ontology in library and information science.

\bibitem{b3}
Zhao, X., Lopez, S., Saikin, S., Hu, X.,  Greenberg, J. (2021). Text to insight: Accelerating organic materials knowledge extraction via deep learning. Proceedings of the Association for Information Science and Technology, 58(1), 558–562.

\bibitem{b4}
Author, A.-B.: Contribution title. In: 9th International Proceedings
on Proceedings, pp. 1--2. Publisher, Location (2010)

\bibitem{b5}
Himanen, L., Geurts, A., Foster, A.,  Rinke, P. (2019). Data-driven materials science: status, challenges, and perspectives. Advanced Science, 6(21), 1900808.

\bibitem{b6}
Kumaraguru, S., Rachuri, S.,  Lechevalier, D. (2014). Faceted classification of manufacturing processes for sustainability performance evaluation. The International Journal of Advanced Manufacturing Technology, 75(9), 1309–1320.

\bibitem{b7}
Zhao, X., Greenberg, J., Meschke, V., Toberer, E., \& Hu, X. (2021). An exploratory analysis: extracting materials science knowledge from unstructured scholarly data. The Electronic Library.

\bibitem{b8}
Simperler, A. and Goldbeck, G. (2021). OntoTrans benefits for industry

\bibitem{b9}
Materials Genome Initiative strategic plan. (2021). Executive Office of the President, National Science and Technology Council, Committee on Technology, Subcommittee on the Materials Genome Initiative.

\bibitem{b10}
Agrawal, A., \& Choudhary, A. (2016). Perspective: Materials informatics and big data: Realization of the “fourth paradigm” of science in materials science. Apl Materials, 4(5), 053208.
\bibitem{b11}
Hey, A., Tansley, S., Tolle, K., \& others (2009). The fourth paradigm: data-intensive scientific discovery. (Vol. 1) Microsoft research Redmond, WA.

\bibitem{OBO}
Obo foundry.
\newblock \url{http://www.obofoundry.org/}.

\bibitem{NCBO}
Ncbo bioportal.
\newblock \url{http://bioportal.bioontology.org/}.

\bibitem{matportal}
\url{https://matportal.org/}

\bibitem{b12}
Bliss, H., \& others (1935). System of bibliographic classification.

\bibitem{b13}
Dewey, M. (1876). A classification and subject index, for cataloguing and arranging the books and pamphlets of a library. Brick row book shop, Incorporated.

\bibitem{b14}
Falcon, A. (2006). Aristotle on causality. 

\bibitem{b15}
Feng, J., Su, B.L., Xia, H., Zhao, S., Gao, C., Wang, L., Ogbeide, O., Feng, J., \& Hasan, T. (2021). Printed aerogels: Chemistry, processing, and applications. Chemical Society Reviews, 50(6), 3842–3888.

\bibitem{b16}
Hawley, W., \& Li, J. (2019). Electrode manufacturing for lithium-ion batteries—Analysis of current and next generation processing. Journal of Energy Storage, 25, 100862.

\bibitem{b17}
Bayerlein, B., Hanke, T., Muth, T., Riedel, J., Schilling, M., Schweizer, C., Skrotzki, B., Todor, A., Moreno Torres, B., Unger, J., \& others (2022). A Perspective on Digital Knowledge Representation in Materials Science and Engineering. Advanced Engineering Materials, 2101176.

\bibitem{b18}
Guerrini, M. (2004). Dewey Decimal Classification and Relative Index, devised by Melvil Dewey, Ed. 22. Bollettino AIB (1992-2011), 44(2), 199–201.

\bibitem{b19}
Ranganathan, S. R. (1933). Colon classification. In Madras Library Association, 1933. 1v.

\bibitem{b20}
Greenberg, J., Zhao, X., Adair, J., Boone, J., \& Hu, X. (2020). HIVE-4-MAT: Advancing the ontology infrastructure for materials science. In Research Conference on Metadata and Semantics Research (pp. 297–307).

\bibitem{b21}
La Barre, K. (2006). The use of faceted analytico-synthetic theory as revealed in the practice of website construction and design. Indiana University.

\bibitem{b22}
Federal Funds for Research and Development: Fiscal Years 2013–15

\bibitem{b23}
Fidel, R. (1984). The case study method: A case study. Library and Information Science Research, 6(3), 273–288.

\bibitem{b24}
Ashino, T. Materials ontology: An infrastructure for exchanging materials information and knowledge. {\em Data Science Journal}. \textbf{9} pp. 54-61 (2010)

\bibitem{b25}
Vet, P., Speel, P. \& Mars, N. The Plinius ontology of ceramic materials. {\em Eleventh European Conference On Artificial Intelligence (ECAI’94) Workshop On Comparison Of Implemented Ontologies}. pp. 8-12 (1994)


\end{thebibliography}
%

\end{document}